\documentclass[superscriptaddress, noeprint, amsmath,amssymb,aps,twocolumn]{revtex4-2}
\usepackage{graphicx}
\usepackage{dcolumn}
\usepackage{xcolor}
\usepackage{bm}
\usepackage{hyperref}
\usepackage{upgreek}

\usepackage{verbatim}

\newcommand{\kb} {\ensuremath{k_{\mathrm{B}}}}
\newcommand{\tb} {\ensuremath{T_{\mathrm{MF}}}}
\newcommand{\tth} {\ensuremath{T_{\Theta}}}
\newcommand{\x} {\ensuremath{x}}
\newcommand{\tbkt} {\ensuremath{T_{\mathrm{BKT}}}}

\newcommand{\lmin} {\ensuremath{\lambda_{\mathrm{min}}}}
\newcommand{\lo} {\ensuremath{\lambda_0}}
\newcommand{\los} {\ensuremath{\lambda_0^{-2}}}
\newcommand{\nsno} {Nd$_{1-x}$Sr$_{x}$NiO$_2$}
\newcommand{\figref}[2]{Fig. \ref{#1}{\color{blue}#2}}

\hypersetup{bookmarksnumbered=true,	unicode=false,	pdfstartview={FitH}, pdfnewwindow=true, colorlinks=true,linkcolor=blue, citecolor=blue, filecolor=blue, urlcolor=blue}

\begin{document}

\title{Evolution of the Superfluid Density in Infinite-Layer Nickelates}

\author{Bai Yang Wang}
\thanks{These authors contributed equally}
\email{bwang87@stanford.edu}
\affiliation{Stanford Institute for Materials and Energy Sciences, SLAC National Accelerator Laboratory,  Menlo Park, CA USA}
\affiliation{Department of Physics, Stanford University, Stanford, CA, USA}

\author{Shannon P. Harvey}
\thanks{These authors contributed equally}
\email{}
\affiliation{Stanford Institute for Materials and Energy Sciences, SLAC National Accelerator Laboratory,  Menlo Park, CA USA}
\affiliation{Department of Applied Physics, Stanford University, Stanford, CA, USA}
 
\author{Kyuho Lee}
\affiliation{Stanford Institute for Materials and Energy Sciences, SLAC National Accelerator Laboratory,  Menlo Park, CA USA}
\affiliation{Department of Physics, Stanford University, Stanford, CA, USA}

\author{Yijun Yu}
\affiliation{Stanford Institute for Materials and Energy Sciences, SLAC National Accelerator Laboratory,  Menlo Park, CA USA}
\affiliation{Department of Applied Physics, Stanford University, Stanford, CA, USA}

\author{Yonghun Lee}
\affiliation{Stanford Institute for Materials and Energy Sciences, SLAC National Accelerator Laboratory,  Menlo Park, CA USA}
\affiliation{Department of Applied Physics, Stanford University, Stanford, CA, USA}

\author{Motoki Osada}
\affiliation{Stanford Institute for Materials and Energy Sciences, SLAC National Accelerator Laboratory,  Menlo Park, CA USA}
\affiliation{Department of Applied Physics, Stanford University, Stanford, CA, USA}

\author{Chaitanya Murthy}
\affiliation{Department of Physics, Stanford University, Stanford, CA, USA}
\affiliation{Department of Physics and Astronomy, University of Rochester, Rochester, NY, USA}

\author{Srinivas Raghu}
\affiliation{Stanford Institute for Materials and Energy Sciences, SLAC National Accelerator Laboratory,  Menlo Park, CA USA}
\affiliation{Department of Physics, Stanford University, Stanford, CA, USA}

\author{Harold Y. Hwang}
\email{hyhwang@stanford.edu}
\affiliation{Stanford Institute for Materials and Energy Sciences, SLAC National Accelerator Laboratory,  Menlo Park, CA USA}
\affiliation{Department of Applied Physics, Stanford University, Stanford, CA, USA}

\date{\today}

\begin{abstract}
 Nickelate superconductors provide a valuable new platform for the study of unconventional superconductivity that is complementary to the cuprates. One of the central puzzles about high-temperature superconductors is what factors determine the scale of their superconducting transition temperature ($T_{\mathrm{c}}$). To address this question for infinite-layer nickelates, we present a systematic mutual inductance study of the superfluid density across the doping-dependent superconducting dome of Nd$_{1-x}$Sr$_{x}$NiO$_2$. We observe a weak superfluid stiffness that exhibits an approximately square-root correlation with $T_{\mathrm{c}}$. We also find a strong interplay between Nd magnetism and the superconducting phase, manifested as a substantial low-temperature suppression of superfluid density. These observations highlight the importance of superconducting phase fluctuations in limiting $T_{\mathrm{c}}$ and unexpectedly strong coupling between the Nd 4$f$ moments and the superfluid.
\end{abstract}

\maketitle

Since the initial report of superconducting \nsno{} thin films \cite{li2019d}, the family of nickelate superconductors has rapidly expanded to include the reduced $\mathrm{Nd_6Ni_5O_{12}}$ \cite{pan2022a} and members of the perovskite Ruddlesden-Popper series, $\mathrm{La_3Ni_2O_7}$ and $\mathrm{La_4Ni_3O_{10}}$ \cite{sun2023,Zhu2024}.  At the same time, the maximum superconducting transition temperature ($T_{\mathrm{c}}$) has been increased via chemical substitution in the infinite-layer nickelates \cite{liner2025}, and exceeds liquid nitrogen temperature in $\mathrm{La_3Ni_2O_7}$, comparable to that of cuprates. This establishes the nickelates as a family of high-temperature superconductors spanning multiple structural and electronic manifestations. From a physics point of view, distinctions between the nickelates and the cuprates have opened up several new perspectives. For one, the nominal $3d^{7.5}$ and $3d^{7.33}$ electronic configurations of Ni in $\mathrm{La_3Ni_2O_7}$ and $\mathrm{La_4Ni_3O_{10}}$ challenge the perception of high-temperature superconductors as doped Mott insulators. The absence of nearby long-range magnetic order in some nickelate superconductors has also raised questions about the relevance of pairing mediated by magnetic fluctuations. Identifying universal characteristics across the cuprate and nickelate families has important implications for developing a deeper understanding of high-temperature superconductivity.

In this context, one essential superconducting property to examine is the superfluid density. In a conventional BCS superconductor, all normal-state carriers condense into the superfluid once the temperature is low enough for Cooper pairs to form. Therefore, the $T_{\mathrm{c}}$ scale is independent of the superfluid density and determined solely by the pairing interaction strength. In the cuprate superconductors, however, a large fraction of the normal-state carriers does not condense, and the superfluid density exhibits a clear positive correlation with $T_{\mathrm{c}}$ \cite{uemura1988}. These observations highlight the crucial role of superconducting phase stiffness in determining the $T_{\mathrm{c}}$ scale and thereby the shape of the superconducting dome \cite{emery1995a}. Such a reduced stiffness could be a generic feature of high-temperature superconductivity and may also be present in the nickelate superconductors. So far, the infinite-layer nickelates have been the most extensively characterized, following progress in synthesis optimization \cite{zeng2020,wei2023,lee2022}. Systematic mapping of their doping phase diagram has indeed revealed a superconducting dome strikingly similar to that of the cuprates, providing an ideal material platform to address this question.

Here we study samples of \nsno{} with $x$ ranging from 0.12 to 0.25, spanning the superconducting dome. The samples are approximately 5 nm thick and are grown using pulsed laser deposition of the perovskite form of the material on (LaAlO$_3$)$_{0.3}($Sr$_2$TaAlO$_6)_{0.7}$ (LSAT) substrates, capped with SrTiO$_3$ (STO), and subsequently reduced chemically to the infinite-layer phase, as described in \cite{lee2020a}. This represents the highest-crystallinity doping series that we have achieved \cite{lee2020a,lee2022}, by managing the lattice mismatch with the substrate. We measure the complex conductance $\sigma_1(T) - i \sigma_2(T)$ of these samples using a two-coil mutual inductance technique \cite{turneaure1998}. Here $\sigma_1$ is the dissipative component of the signal, which exhibits a peak at the superconducting transition and serves as a measure of the transition width and sample homogeneity \cite{bozovic2016}. $\sigma_2$ is the inductive component of the signal, from which we extract the in-plane magnetic penetration depth, $\lambda$. This penetration depth is related to the superfluid density through the relation $\lambda(T)=\sqrt{\frac{m^*}{4 \mu_0 n_s(T) e^2}}$, where $\mu_0$ is the vacuum permeability, $e$ is the electron charge, $n_s$ is the superfluid density, and $m^*$ is the electronic effective mass. Based on this, we can effectively probe the superfluid stiffness, $J_s\propto n_s/m^* \propto 1/\lambda^2$ \cite{prozorov2006, hardy2002}. The measurements are performed at 30-60 kHz and down to 150 mK, and more details about the apparatus and measurement procedures are described in \cite{harvey2025}.

\begin{figure}
	\includegraphics[width=3.425in]{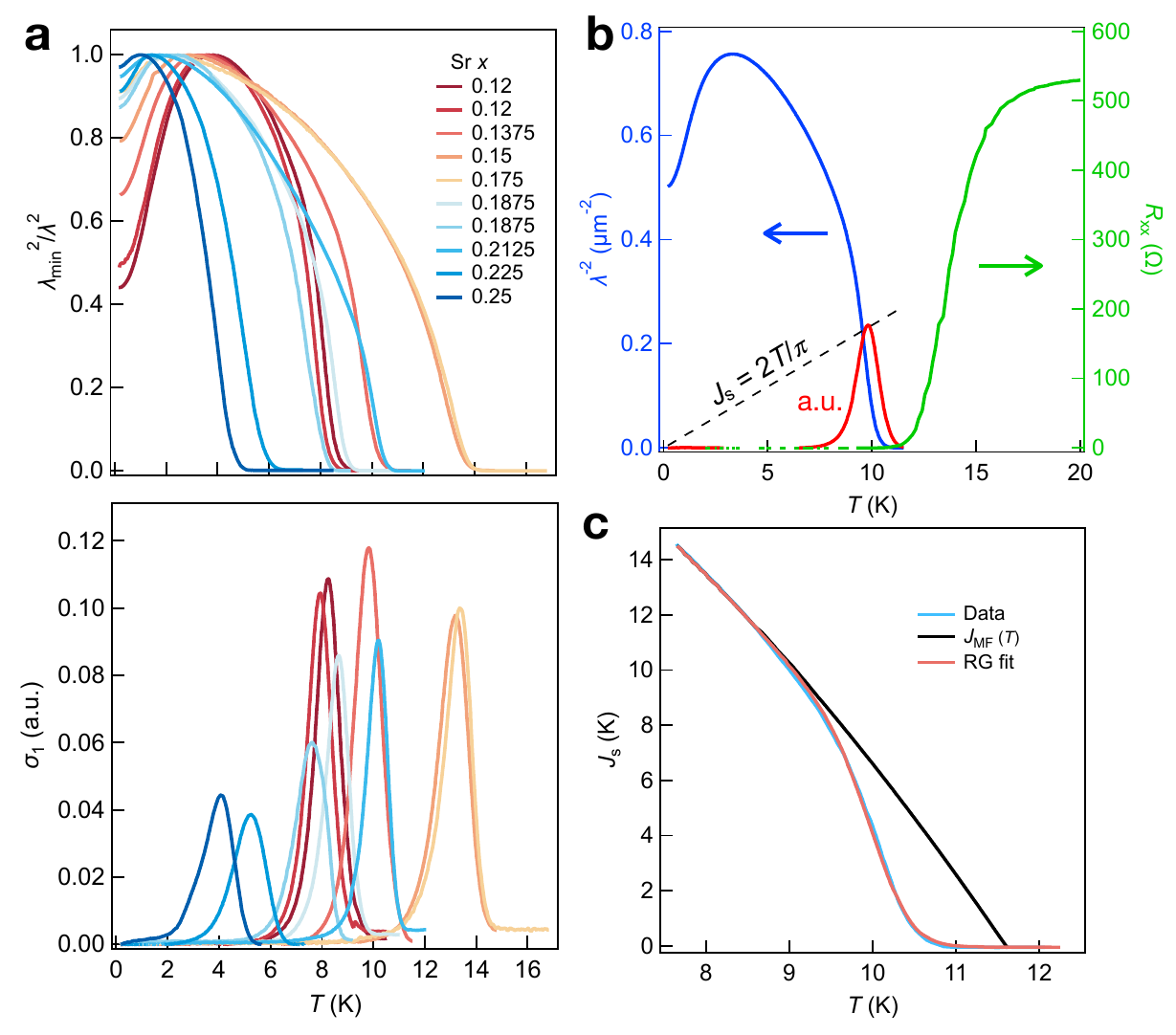}
	\caption{(a) The full temperature dependence of the superfluid density (upper panel, normalized to its maximum value) and the dissipative signal (lower panel) for 10 samples spanning the superconducting dome. The legend describes the Sr doping value. 
 (b) The superfluid density (solid blue line, left axis), dissipative signal (solid red line, arbitrary units), and resistance (solid green line, right axis) as a function of temperature for for a representative Nd$_{0.8625}$Sr$_{0.1375}$NiO$_2$ sample. The dashed black line marks where superfluid stiffness equals thermal fluctuations in terms of energy scale.
 (c) Temperature dependence of superfluid stiffness near \tbkt{}, with the mean field stiffness, data, and the fit to the stiffness using the renormalization group equations for a representative Nd$_{0.7875}$Sr$_{0.2125}$NiO$_2$ sample.}
	\label{fig1}
\end{figure}

In \figref{fig1}{(a)}, the normalized superfluid density and dissipative signal are presented as functions of temperature for each of the \nsno{} samples. Each transition is well-defined and complete far above zero temperature. The full width at half maximum of the transition in the dissipative channel varies within 2-3 K, indicating macroscale sample homogeneity. This is, by comparison, difficult to achieve in cuprates near the edges of the doping-dependent dome \cite{broun2007}, and reflects the improved sample quality achieved through the use of the LSAT substrate, which better stabilizes growth and reduction conditions \cite{lee2022}. The improved crystallinity also reveals several important features in the temperature dependence of the superfluid density. In \figref{fig1}{(b)}, we plot the mutual inductance data and the corresponding transport characterization for a representative sample with $x$ = 0.1375. At temperatures below about 4 K, there is a clear suppression of the superfluid density as the temperature is reduced, which we attribute to the interplay between the superconductivity and the Nd 4$f$ moments. Near the superconducting transition, there is also an accelerated decrease in the superfluid density as the temperature increases, marking the Berezinskii-Kosterlitz-Thouless (BKT) transition.

This BKT transition describes the unbinding of vortex-antivortex pairs due to thermal fluctuations, manifested as the observed loss of superfluid density \cite{kosterlitz1973}. It occurs when the thermal fluctuation energy matches the superfluid stiffness, as described by $J_s(\tbkt) = 2 \tbkt/\pi$. Examination of \figref{fig1}{(b)} shows that at the temperature \tbkt{} where the superfluid drop occurs, $\ell\approx d$, the film thickness. This confirms that superconductivity is coherent throughout the film thickness, i.e., $\xi_c(T\approx T_{\mathrm{c}})\geq d/\sqrt{\pi}$. This behavior is similar to that observed in $\mathrm{La}_{1-x}\mathrm{Sr}_x\mathrm{CuO_4}$ (LSCO) and $\mathrm{YBa_2Cu_3O}_{7-x}$ (YBCO), but in contrast to Bi$_2$Sr$_2$Ca$_3$CuO$_{8+\delta}$, where the relevant thickness is one to two monolayers \cite{hetel2007, lemberger2010, yong2012}. As shown in \figref{fig1}{(c)} for another representative sample with $\x=0.2125$ exhibiting a clear BKT transition, we can also quantitatively capture the BKT transition by fitting the data with a renormalization-group-based model \cite{benfatto2009}. Details and additional fits to other data sets are discussed in the Supplementary Material \cite{SupplementalMaterial}. The fit includes two free parameters: the vortex core energy $\mu$ and the inhomogeneity $\delta$, the latter of which consistently tracks the transition width in $\sigma_1$ (see Supplementary Material \cite{SupplementalMaterial} and references \cite{turneaure1996,ganguly2015,wang2022h} therein). These observations again confirm the sample quality and the completeness of the superconducting transition.

\begin{figure}
	\includegraphics[width=3.425 in]{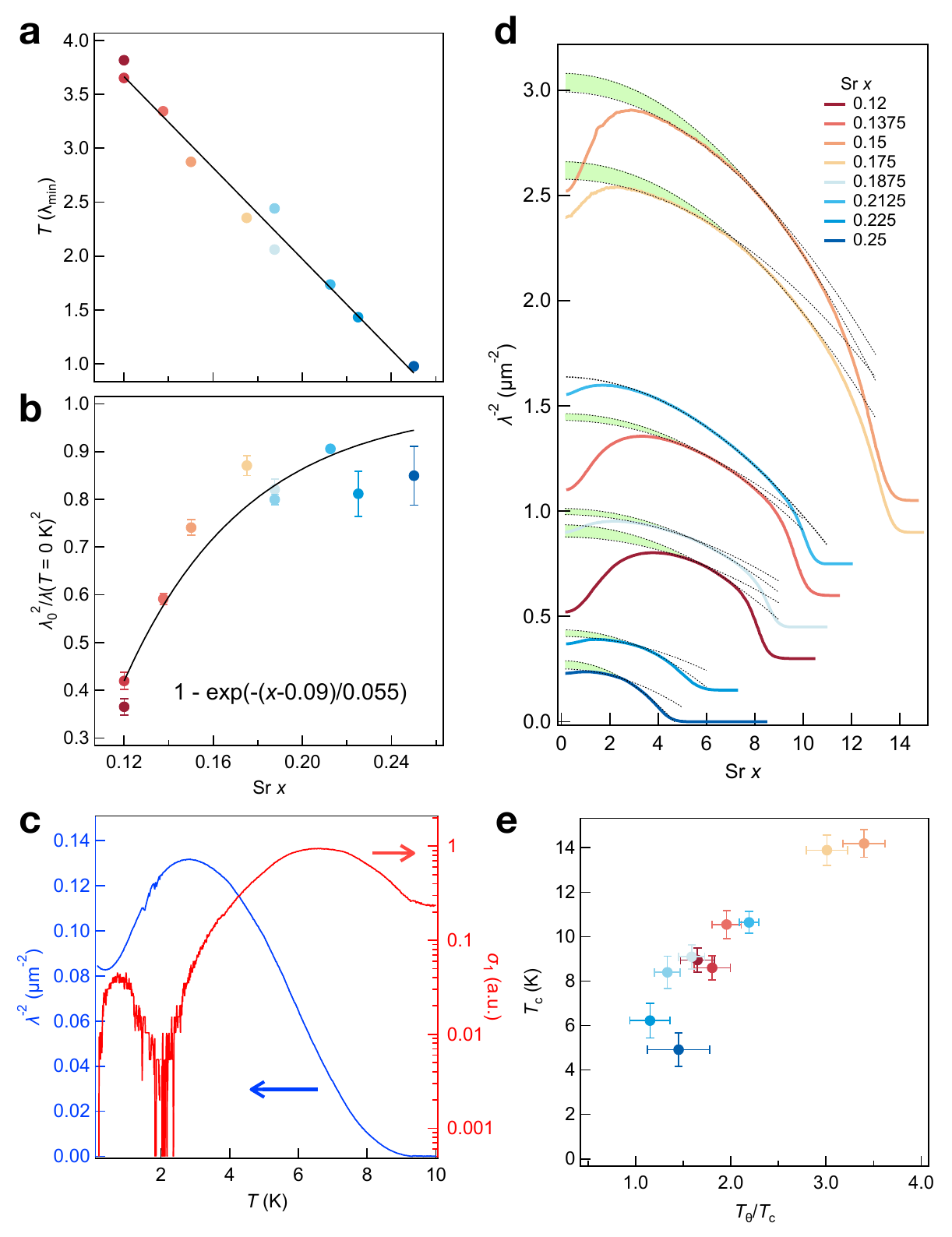}
	\caption{ (a) The temperature at which the penetration depth reaches its minimum value, $T$($\lambda=\lmin$), as a function of Sr doping $\x$, showing an approximately linear dependence, with fit line (black) $T(\lambda=\lmin)=21(0.29-x)$. (b) The ratio of measured superfluid density remaining at base temperature $\lambda(T = 0 \ \mathrm{K})^{-2}$ to its estimated maximum value in the absence of magnetic effects $\lambda_0^{-2}$, as a function of Sr doping. The solid black fit line follows $\lo^2/\lmin^2=1- \exp(-(x-0.09)/0.055)$.  (c) The superfluid density and the dissipative signal of an underdoped Nd$_{0.88}$Sr$_{0.12}$NiO$_2$ sample, where the maximum superfluid suppression is observed. (d) The full temperature dependence of the superfluid density proxy $\lambda_0^{-2}$ for each doping. For each sample, a pair of dashed black lines shows our quadratic fits obtained for the temperature ranges near $T(\lambda=\lmin)$ and the onset of the BKT superfluid drop, with the green shade illustrating the range of variation of our estimate. (e) $T_{\mathrm{c}}$ as a function of \tth{}, with the color of each marker representing its corresponding Sr doping, following the coloring scheme in panel (a) and (b).}
	\label{fig2}
\end{figure}

The low-temperature decrease in superfluid density is less commonly observed. Qualitatively similar behavior has been reported in cuprate and iron pnictide materials containing Nd and is attributed to the paramagnetism of the  Nd$^{3+}$ ions \cite{cooper1996, prozorov2000a, martin2009,alff1999}. Indeed, such effects are not seen in La- or Pr-nickelates, where La and Pr have nonmagnetic ground states \cite{prozorov2000a, martin2009, harvey2025, wang2022h}. However, the extent of the observed superfluid suppression in our samples is substantially larger, by more than an order of magnitude. Moreover, our data do not show a Curie-Weiss-like temperature dependence, and some even exhibit a small recovery of the superfluid density at the lowest temperatures. These features are reminiscent of the (much smaller) superfluid modulations observed in some iron pnictides and rare-earth nickel borocarbides \cite{chia2005, gasparov2009}, and suggest an interplay between magnetic ordering of the Nd$^{3+}$ moments and the superconducting phase that goes beyond simple paramagnetism.

If we approximate the transition temperature of this putative magnetic order by the temperature of the minimum measured penetration depth, \lmin{}, the associated phase boundary is shown in \figref{fig2}{(a)}. It exhibits a linear doping dependence, extrapolating to zero at $x=0.29$. Such a linear dependence closely resembles that observed in magnetically ordered systems diluted with nonmagnetic substitutes \cite{keimer1992, lora-serrano2009}. In this case, the dilution follows the doping by substituting non-magnetic Sr ions for Nd ions. Correspondingly, as shown in \figref{fig2}{(b)}, the extent of this superfluid reduction is also clearly doping dependent, with the superfluid density at base temperature rapidly approaching the value estimated in the absence of magnetic effects as the doping approaches $x=0.29$ as discussed below.

The effect of this interplay is also visible in the dissipative channel. As shown in \figref{fig2}{(c)}, a quasiparticle dissipation peak develops simultaneously with the suppression of superfluid density at low temperatures, indicating a loss of superfluid density to normal-state quasiparticles. The lack of magnetic anisotropy in the external-field suppression of the superfluid density (see the Supplementary Material \cite{SupplementalMaterial} and references \cite{turneaure1996,ganguly2015,wang2022h} therein) points to a more fundamental coupling beyond the magnetic susceptibility modulations observed previously in angular magnetoresistance and upper critical field measurements of the Nd-nickelates \cite{wang2022h}. Indeed, our observations indicate long-range magnetic order of the Nd\textsuperscript{3+} magnetic moments at the temperature scale of 3 - 4 K, extending to the overdoped end of the superconducting dome, and consistent with the upper critical field analysis in ref.\cite{wang2021g}. The loss of superfluid density is most substantial on the underdoped side of the dome, and suggests scenarios such as the loss of Fermi surface due to Brillouin zone folding upon antiferromagnetic ordering. Despite experimental challenges, further studies of this low-temperature regime may be quite illuminating for the interactions governing the superconducting state.



To estimate the superfluid density near zero temperature in the absence of Nd magnetism, we perform a quadratic fit that closely captures the temperature dependence of the superfluid density in La- and Pr-nickelates, where rare-earth magnetism is not a factor \cite{harvey2025,wang2022h}: $\lambda^{-2} (T) = \lambda^{-2}_0 (1-(T/\tb)^2)$. Here \tb{} represents the expected transition temperature from mean-field effects, and $\lambda_0$ is the zero-temperature penetration depth. We limit the fit to the temperature range between approximately $T(\lambda=\lmin)$ and the onset of the BKT superfluid drop.

In \figref{fig2}{(d)}, we illustrate how our estimation of the non-magnetic zero-temperature superfluid density varies with the chosen fitting range. For each solid-colored line representing a set of measured data, two black dashed traces indicate the quadratic fits obtained for temperature ranges near $T(\lambda=\lmin)$ and near the onset of BKT superfluid drop, respectively. Together, these two fits provide the lower and upper bounds of our estimate, the average of which serves as our final estimation of the non-magnetic zero-temperature superfluid density, $\lo^{-2}$.

\begin{figure*}
	\includegraphics[width=6.45 in]{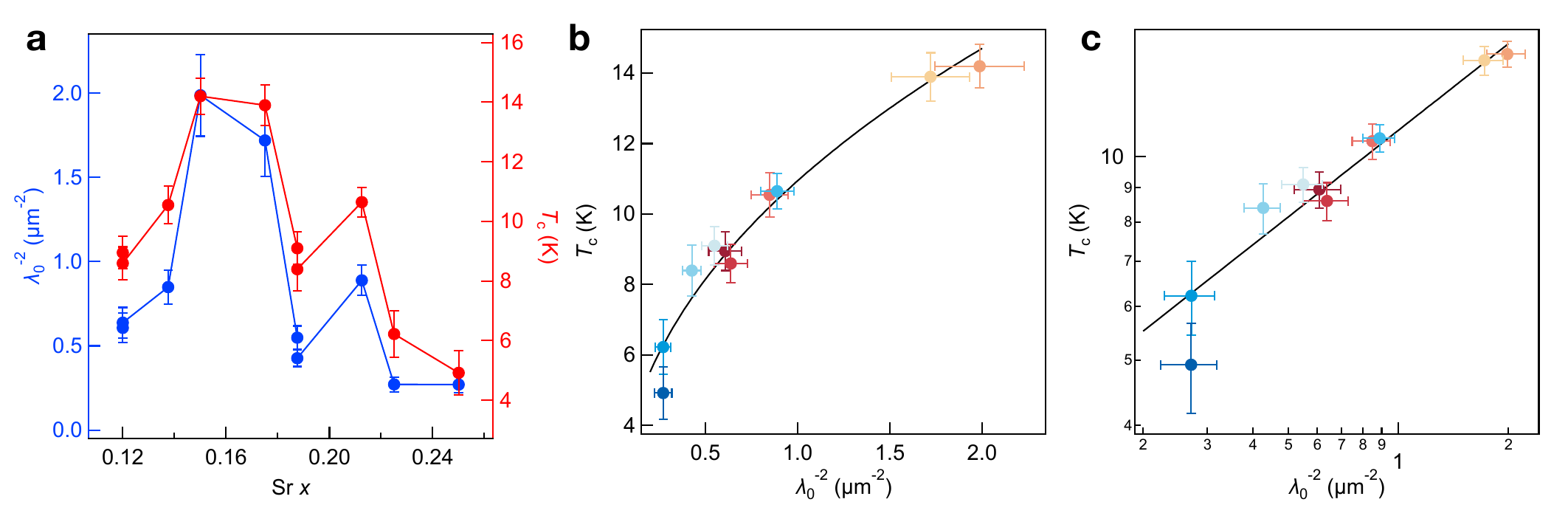}
	\caption{(a) \los{} (left, blue) and $T_{\mathrm{c}}$ (right, red) measured as a function of doping across the superconducting dome. 
 (b) Linear plot of $T_{\mathrm{c}}$ against \los{}. A power law fit to the data is shown as a solid black line: $T_{\mathrm{c}} =(10.9 \pm 0.2)  \left( \los \right) ^{0.43 \pm 0.03}$. The color of each marker represents its corresponding Sr doping, following the coloring scheme in \figref{fig1}{} and \figref{fig2}{}.
 (c) Log-log plot of $T_{\mathrm{c}}$ against \los{}. The same fit line in panel (b) is shown again as a solid black line.}
	\label{fig3}
\end{figure*}

Our estimation of \lo{} allows us to examine the evolution of the superfluid stiffness. This energy term, which characterizes the rigidity of the superconducting phase, is defined as $J_s(T)=\frac{\hbar^2}{4 \mu_0 e^2 \kb}\frac{\ell}{\lambda^2(T)}$, where $\ell$ represents the length scale along the $c$-axis over which the material is phase-locked \cite{benfatto2009, benfatto2012}. This length scale is the larger of the $c$-axis lattice spacing ($c$ = 0.334 nm) and the $c$-axis coherence length $\sqrt{\pi} \xi_c(0)$. To be conservative, we assume $\ell = d$, where $d$ is the film thickness. Therefore, we can estimate an upper bound for the approximate temperature above which superconducting phase coherence is destroyed by thermal fluctuations as $\tth= A J_s(0) \times \pi/2$, where A is a constant of order 1 \cite{emery1995a}. Setting $A=0.9$, appropriate for quasi-two-dimensional superconductors, \tth{} ranges from 1 to 4 $T_{\mathrm{c}}$ across the superconducting dome. Here, $T_{\mathrm{c}}$ is defined as the temperature at which superfluid density reaches 1\% of its maximum value. As shown in \figref{fig2}{(e)}, the samples with doping spanning the superconducting dome fall on a single curve where $T_{\mathrm{c}}$ drops rapidly as $\tth/T_{\mathrm{c}}$ approaches 1. The weak superfluid stiffness, together with the clear correlation between $T_{\mathrm{c}}$ and $\tth/T_{\mathrm{c}}$, underscores the importance of phase coherence in stabilizing the superconducting phase in infinite-layer nickelates, particularly near the dome edge. 

 Next we examine the doping dependence of \los{} and its correlation with $T_{\mathrm{c}}$. As shown in \figref{fig3}{(a)}, \los{} and $T_{\mathrm{c}}$ closely track each other across doping. In \figref{fig3}{(b),(c)}, we plot these two quantities on linear and log-log scales, with the error in $T_{\mathrm{c}}$ given by the transition width and the error in \los{} representing the combined uncertainty from the coil geometry and from the fit used to remove magnetic effects. We find a clear correlation between $T_{\mathrm{c}}$ and \los{}, to which we perform a power-law fit. The best fit yields $T_{\mathrm{c}} =(10.9 \pm 0.2)  \left( \los \right) ^{0.43 \pm 0.03}$, where $T_{\mathrm{c}}$ is in kelvin and \lo{} in micrometers. Notably, samples across the entire superconducting dome follow the same trend, highlighting an intriguing discrepancy between the normal-state carrier density and the superfluid density --- similar to that observed in the cuprates. To account for potential uncertainties in defining $T_{\mathrm{c}}$, we further fit \los{} versus \tb{} and \tbkt{} (see the Supplementary Material \cite{SupplementalMaterial} and references \cite{turneaure1996,ganguly2015,wang2022h} therein) and find that the scaling behavior is minimally affected.

\begin{figure}
	\includegraphics[width=3.425 in]{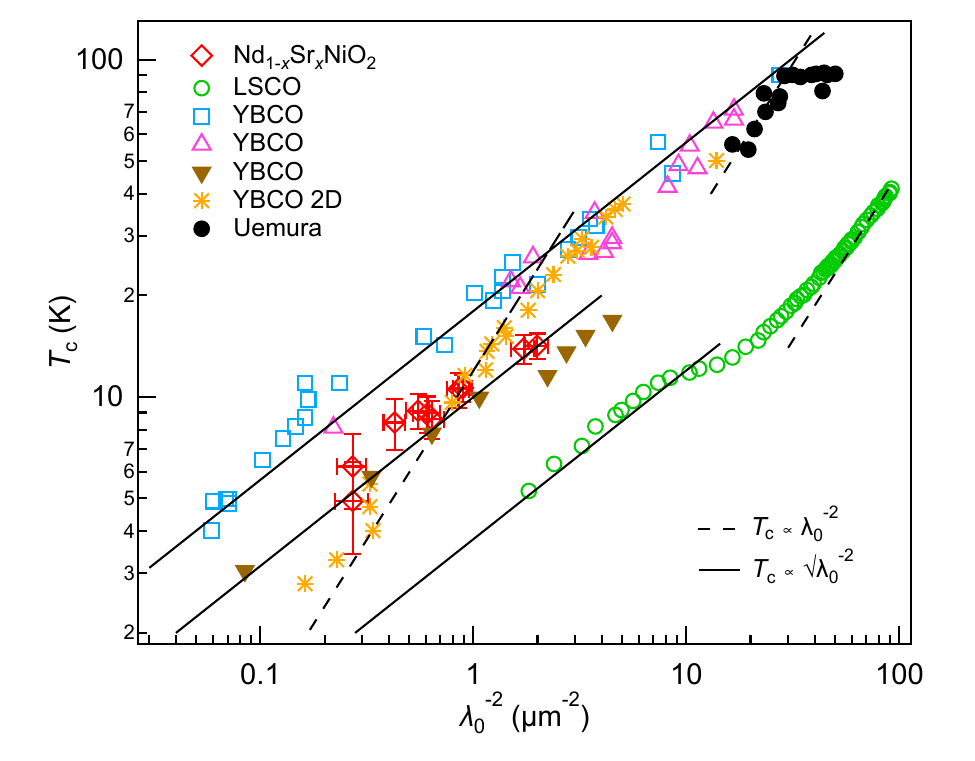}
	\caption{a) $T_{\mathrm{c}}$ as a function of $\lambda_0^{-2}$ across doping for \nsno{} and representative cuprate superconductors: LSCO and YBCO. Solid symbols correspond to bulk samples and open symbols correspond to thin film samples. Black lines are guides to the eye, with the dashed ones corresponding to $T_{\mathrm{c}} \propto \lambda_0^{-2}$ and solid ones to $T_{\mathrm{c}} \propto \sqrt{\lambda_0^{-2}}$. Red diamonds: our \nsno{} data; Green circles: \cite{bozovic2016}; Blue squares: \cite{zuev2005}; Pink triangles: \cite{hetel2007}; Brown triangles: \cite{broun2007}; Orange stars: \cite{hetel2007}; Black circles: \cite{uemura1989}, converted following \cite{zuev2005}.}
	\label{fig4}
\end{figure}

Useful insight can be gained by comparing the data from infinite-layer nickelates with that from other unconventional superconductors where $T_{\mathrm{c}}$ can be tuned. In \figref{fig4}{}, we present a series of measurements of $T_{\mathrm{c}}$ versus in-plane \los{} for LSCO and YBCO, in both bulk and thin-film forms, alongside measurements of \nsno{}. Two power-law scalings can be identified. One shows a linear correlation between $T_{\mathrm{c}}$ and the superfluid density, indicated by dashed black lines ($T_{\mathrm{c}} \propto \lambda_0^{-2}$); the other shows a square-root dependence, indicated by solid black lines ($T_{\mathrm{c}} \propto \sqrt{\lambda_0^{-2}}$). For the cuprates, the linear scaling is mostly observed for dopings away from the edges of the superconducting dome --- such as in the original $\mu$SR data of underdoped bulk YBCO (solid black circles) and in the extensive mutual-inductance data of overdoped LSCO thin films (open green circles). In contrast, the square-root scaling emerges near the low-$T_{\mathrm{c}}$, low-superfluid-density corner of \figref{fig4}{}, as doping approaches either the underdoped (open blue squares and pink triangles) or overdoped (open green circles) edges of the dome.

We also plot one exception: an ultrathin, 2-unit-cell-thick YBCO thin film (orange stars), where $T_{\mathrm{c}}$ drops considerably faster, following a linear dependence on superfluid density. This deviation has been attributed to dimensionality-dependent scaling behavior near a quantum critical point in the two-dimensional limit. Our \nsno{} data reside in the low-$T_{\mathrm{c}}$ and low-superfluid-density corner, and show a striking similarity to the cuprates in the three-dimensional limit --- both in the value of $T_{\mathrm{c}}$ at a given \los{} and in the square-root scaling behavior. This is also consistent with previous estimates of $\xi_c(0) < 1$ nm $< d$ for \nsno{} \cite{wang2021g}.

We next consider potential explanations for the observed correlation between $T_{\mathrm{c}}$ and the superfluid density. A three-dimensional quantum critical point has been proposed to lead to square-root scaling, which may explain such behaviors in LSCO and YBCO \cite{hetel2007, lemberger2011}. However, this scaling is observed for \nsno{} across the entire superconducting dome. This distinction raises the possibility that low superfluid density alone — independent of fine-tuned quantum criticality — can generically produce square-root scaling in unconventional superconductors.

In nodal superconductors, disorder has a large impact on both $T_{\mathrm{c}}$ and the superfluid density and can therefore produce an indirect correlation between them \cite{hirschfeld1993, franz1997, sun1995a, lee-hone2017}. Indeed, characterizations of the gap structure in infinite-layer nickelates \cite{harvey2025,chow2022,cheng2024} so far are consistent with a nodal structure, which could allow disorder to play a dominant role \cite{ranna2025}, especially when $T_{\mathrm{c}}$ is relatively low. Nonetheless, the small \tth{} values across the dome, the rapid collapse of $T_{\mathrm{c}}$ as $T_{\mathrm{c}}/\tth$ approaches unity, and the clear positive correlation between $T_{\mathrm{c}}$ and the superfluid density all strongly suggest an integral role of phase fluctuations in explaining this scaling. The similar relationship between $T_{\mathrm{c}}$ and the superfluid density observed in both nickelate and cuprate superconductors suggests that the loss of phase coherence may have comparable importance for superconductivity in both material systems, and that there remains considerable room for further $T_{\mathrm{c}}$ enhancement in infinite-layer nickelates.


\

\acknowledgements{\textcolor{blue}{} We thank Steve Kivelson, Yue Yu and Xinyang Zhang for useful discussions. We thank Jiachen Yu for providing the aluminum thin film for leakage measurements. This work was supported by the US Department of Energy, Office of Basic Energy Sciences, Division of Materials Sciences and Engineering (contract no. DE-AC02-76SF00515). Additional support was provided by Gordon and Betty Moore Foundation’s Emergent Phenomena in Quantum Systems Initiative (grant No. GBMF9072, synthesis equipment), and the Kavli Foundation, Klaus Tschira Stiftung, and Kevin Wells (aspects of the mutual inductance probe).}
\\
\\

\section*{Author Contributions}
B.Y.W., S.P.H., and H.Y.H. conceived the project. K.L., Y.L., and M.O. synthesized the nickelate samples. B.Y.W., S.P.H., and Y.Y. performed the mutual inductance and complementary transport measurements. B.Y.W., S.P.H., C.M., S.R., and H.Y.H. analyzed and interpreted the data. All authors contributed to the manuscript writing.

\bibliography{ref1}
\end{document}